\documentclass[12pt,pra,aps,amssymb,amsfonts,amsmath,tightenlines]{revtex4}
\usepackage{dsfont}
\usepackage{graphicx}
\usepackage{amssymb,amsfonts,amsthm}
\usepackage{color}
\usepackage{verbatim}
\usepackage[normalem]{ulem}
\usepackage{enumerate}
\usepackage{mathrsfs}
\usepackage{bm,float}
\usepackage{amsmath}

\usepackage{textgreek}
\usepackage{lgreek}

\usepackage{caption}
\usepackage{nicefrac}

\usepackage{tikz}
\usetikzlibrary{shapes.geometric}
\begin{document}
\title{Brief Synopsis of the Scientific Career of T. R. Hurd}
\author{Matheus~R.~Grasselli}
\affiliation{\vspace{-.4cm}Department of Mathematics and Statistics, McMaster University, Hamilton, Ontario L8S\,4K1, Canada\\ grassel@mcmaster.ca}
\author{Lane~P.~Hughston}
\affiliation{\vspace{-.4cm}Department of Computing, Goldsmiths University of London\\ New Cross, London SE14\,6NW, United Kingdom\\L.Hughston@gold.ac.uk \vspace{.2cm}}

\begin{abstract}
\noindent 
As an introduction to a Special Issue of International Journal of Theoretical and Applied Finance in Honour of the Memory of Thomas Robert Hurd we present a brief synopsis of Tom Hurd's scientific career and a bibliography of his publications.  
\vspace{-0.3cm}
\\
\end{abstract}

\maketitle


\noindent In this special issue of IJTAF we commemorate the scientific career of our friend and colleague Tom Hurd. In a life cut off too soon he accomplished much and we are mindful of the fact that a brief synopsis of his career cannot possibly capture a sense of the totality of his contribution and the tremendous impact he had on all his areas of research. Nonetheless, since what we have to say will undoubtedly be of interest to his friends, family, collaborators and students, let us do our best by way of an introduction to recall some of the details. 

Thomas Robert Hurd (always ``Tom" to his friends) was born on 13 October 1956, one of four siblings. He grew up in Ottawa and obtained a BSc in Mathematics and Physics at Queen's University, Kingston, Ontario. In the autumn of 1978 he entered the University of Oxford as a research student in the burgeoning, highly active research group of Roger Penrose at the Mathematical Institute, Oxford. Figure 1 shows a snapshot of Tom from his Oxford days as a research student. Those of you who are familiar with Oxford will know that students (as well as academic members of staff) typically have a kind of dual allegiance within the University, to a department on the one hand (The Mathematical Institute, in Tom's case) and on the other hand to one of Oxford's many colleges (Trinity College, in Tom's case). During his first year Tom's supervisor was S.~T.~(Florence) Tsou. Florence left Oxford for CERN at the end of Tom's first year, at which point Lane Hughston took over as first supervisor, with Roger Penrose acting as second supervisor. In fact, Roger Penrose was generous with his time and would see all the students in his group on a regular basis to discuss their research. Throughout the year he would hold a sort of informal seminar in his office on Fridays that the whole group would attend where over many hours they would discuss and present the latest ideas on which they were working. There is no doubt that Tom was deeply impressed by the Penrose research ethos and indeed it was in that setting that Tom had a first taste of real research and the flow of ideas swirling around the development of a new scientific subject. The subject in question  in this case was Penrose's ``twistor theory," which with its potential for applications to gravitation and relativistic quantum theory using modern ideas from complex analysis and algebraic geometry was very exciting. 

\begin{figure}[h!]
 \centering
  \includegraphics[width=\textwidth]{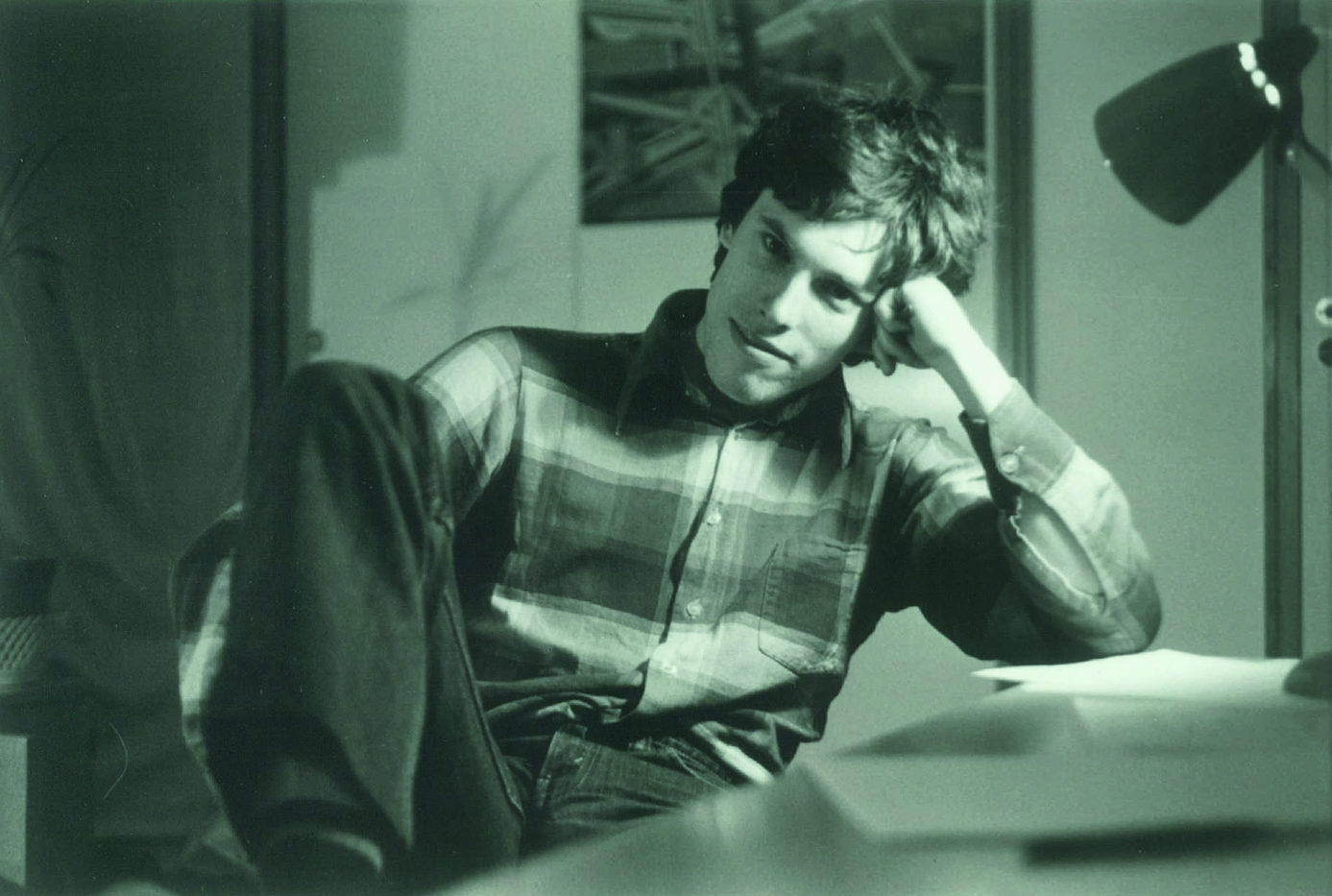}
 \caption{Tom Hurd in Oxford circa 1980.}
 \end{figure}
\vspace{0.5cm} 

The highly cohesive Penrose group developed a kind of informal in-house publication called ``Twistor Newsletter" which would appear in regular issues, and this is where one finds Tom's first output, ``Massless  fields based on the twisted cubic" (Eastwood {\em et al.}~1979).
Many other Twistor Newsletter articles would follow (Hughston \& Hurd 1979, 1980a, 1980b, 1980c, 1981a, 1981b, 1981c, 1981d) leading up to an important publication in Proceedings of the Royal Society on the treatment of massive fields in twistor theory (Hughston \& Hurd 1981e) and then in 1982 to Tom's D.~Phil.~thesis, ``Conformal Geometry and its Applications to Relativistic Quantum Theory" (Hurd 1982).

On the completion of his doctorate Tom obtained a prestigious two-year Weir Junior Research Fellowship at University College, Oxford, where he continued work on applications of twistor theory to relativistic quantum theory (Hughston Hurd 1983a, 1983b, 1983c, Hurd 1984)
and also undertook a series of investigations on the use of twistor methods in relativistic cosmology (Hurd 1983,  Hurd 1985) supported in 1982-1983 by an NSERC Research Assistantship and in 1983-1984 by an SERC Fellowship.

In 1984 Tom moved to University of British Columbia, first as a postdoc and then in 1985 as an Assistant Professor. During his years at UBC Tom's interests in relativistic quantum theory developed in a new direction when he embarked on program to give a rigorous demonstration of the renormalizability of quantum electrodynamics (Hurd 1988a, 1988b, 1983c), leading to a notable book in the Springer Lecture Notes in Physics series (Feldman, Hurd, Rosen \& Wright 1988) and two articles in Communications in Mathematical Physics (Hurd 1989a, 1989b).   

\vspace{0.5cm} 
In 1992 Tom moved to McMaster University in Hamilton, Ontario, as an Assistant Professor. This was not long after he had struck up a collaboration with J.~Dimock leading to a series of important publications concerning the use of renormalization group methods in quantum electrodynamics and in statistical physics (Dimock \& Hurd 1991, 1992a, 1992b, 1993).  He was granted tenure at McMaster in 1992. Then on the back of impressive further advances in these areas (Brydges {\em et al.}~1994a, 1994b, 1995, 1998a, 1998b, Faria da Veiga {\em et al.}~1995, 1996, Hurd 1995, Dimock \& Hurd 2000), now involving additional collaborators including 
D.~Bridges, D.~H.~U.~Marchetti, and P.~A.~Faria da Veiga, Tom was promoted to Professor at McMaster in 1996. 

Perhaps it was inevitable with the end of the century approaching that Tom would once again set out in a new direction, and so it was that he moved decisively into financial mathematics and its applications. Here his contributions were numerous, and he was also in a position to play an influential role in the development of the subject, both at his own university and in the financial mathematics community at large. In 1999 he founded PhiMAC, the highly successful Financial Mathematics Laboratory at McMaster University. His research interests in financial mathematics were broad and included (a) pricing, hedging and optimal portfolio selection in incomplete markets, (b) L\'evy processes in finance, (c) interest rate models, and (d) credit risk modelling, as well as sundry other topics. His output was prolific, again involving a host of coauthors (Choulli \& Hurd 2001, Boland {\em et al.}~2001, Hurd 2004, 2009, 2010, Grasselli \& Hurd 2005, 2007, Yang {\em et al.}~2006, Hurd \& Kuznetsov 2007, 2008, 2009, Hurd Kuznetsov 2008, Ait-Sahalia {\em et al.}~2009, Hurd \& Zhou 2010, Arasaratnam {\em et al.}~2010, Chuang Yi {\em et al.}~2011, Hurd {\em et al.}~2016, Jevtic \& Hurd 2017, Santos-Diaz {\em et al.}~2018, Rigobon {\em et al.}~2022). In 2004 he organized an important workshop in Banff on Semimartingales in Finance. Alongside 
M.~R.~Grasselli, he was lead organizer of the 2010 Fields Institute Six-Month Thematic Program on Mathematical Finance, held at the Fields Institute (Grasselli \& Hurd 2011). And in the same year, in collaboration with S.~Jaimungal, he was lead organizer of the Sixth World Congress of the Bachelier Finance Society, also in Toronto. His pivotal roles in these successful events illustrate the speed with which Tom took on the status of a highly influential leader in the financial mathematics community. 

But more was to follow. The financial crisis of 2007-2008 signaled another important shift in Tom's research agenda, leading him into the area of systemic risk in the global financial system. This was an area that had previously been somewhat neglected by finance theorists and was now urgently in need to a fresh and robust approach with new mathematical tools. This work brought Tom into contact with new collaborators (Hurd \& Gleeson 2013, Gleeson {\em et al.}~2013, Ait-Sahalia \& Hurd 2016, Hurd {\em et al.}~2016, 2017, Hurd 2017, 2018, 2023, Feinstein \& Hurd 2023, Chehaitli {\em et al.}~2024) and ultimately led Tom to write his book ``Contagion! Systemic Risk in Financial Networks" (Hurd 2016), published in the SpringerBriefs in Quantitative Finance series. 

The arrival of the Covid-19 pandemic gave rise to an unexpected final development in Tom's portfolio of research activities, with the realization that, with some adjustments, the models that he had developed for financial contagion could be applied to the spread of infectious diseases (Hurd 2021a, Hurd 2021b, Pang {\em et al.}~2021). One can only imagine what other directions his research might have taken him had he not been struck down by glioblastoma, leading to his untimely death on 28 April 2022.

Throughout his career at McMaster, Tom acted as an outstanding mentor of young mathematicians, and it would not be amiss here to mention by name his seven PhD students Zhuowei Zhou, Chuang Yi, Ping Wu, Quentin Shao, Philippe Deprez, Mohammed Oozeer, and H. Chehaitli; and we also mention the names of his fifteen postdocs, Weijie Pang, Tuan Quoc Tran, Petar Jevtic, Huibin Cheng, Klass Schulze, Cesar Gomez, Alexey Kuznetsov, Matheus Grasselli, Tahir Choulli, Jeffrey Boland, Maria Gorina, Alexander Teplyaev, Brian C. Hall, Albertus Hof, and Domingos Marchetti. Additionally, there were numerous MSc students and Masters in Financial Mathematics students at McMaster who undertook projects and theses under Tom's supervision. So many names! So many mathematicians! Yet, they all benefitted from Tom's generous tutelage and mentorship. 

Much the same can be said for the ease with which he established good working relations with colleagues both at his university and in his subject areas across the globe. We recall the many years over which he chaired the organizing committee of the Fields Institute Quantitative Finance Seminar series at the Fields Institute, Toronto, and the key role he played in his many years as a Managing Editor at IJTAF. 
We recall his long list of successful grant applications and the postdoctoral positions and various activities these grants supported. We recall his numerous academic visits abroad to spend time with colleagues who were always delighted to play host to Tom and his wife Rita Bertoldi on their trips. So much more could be said. The papers gathered in this issue of IJTAF offer, we think, a fit tribute to Tom Hurd's wide interests and multiple contributions in financial mathematics and its applications: we commend them to the reader in commemoration of his fine career.


\vspace{1cm} 


\noindent {\bf Bibliography of T. R. Hurd's scientific works}



\end{document}